\documentclass[%
reprint,
amsmath,amssymb,
aps,
14pt]{revtex4-1}

\usepackage{graphicx}
\usepackage{dcolumn}
\usepackage{bm}
\usepackage{hyperref}

\begin{document}

\title{Element sensitive reconstruction of nanostructured surfaces with finite elements and grazing incidence soft X-ray fluorescence}

\author{V. Soltwisch}
\email{Victor.Soltwisch@ptb.de}
\affiliation{Physikalisch-Technische Bundesanstalt (PTB), 
Abbestr. 2-12, 10587 Berlin, Germany}

\author{P. Hoenicke}
\affiliation{Physikalisch-Technische Bundesanstalt (PTB), 
Abbestr. 2-12, 10587 Berlin, Germany}

\author{Y. Kayser}
\affiliation{Physikalisch-Technische Bundesanstalt (PTB), 
Abbestr. 2-12, 10587 Berlin, Germany}

\author{J. Eilbracht}
\affiliation{Physikalisch-Technische Bundesanstalt (PTB), 
Abbestr. 2-12, 10587 Berlin, Germany}

\author{J. Probst}
\affiliation{Helmholtz-Zentrum Berlin (HZB), Albert-Einstein-Str. 15, 12489 Berlin, Germany}

\author{F. Scholze}
\affiliation{Physikalisch-Technische Bundesanstalt (PTB), 
Abbestr. 2-12, 10587 Berlin, Germany}

\author{B. Beckhoff}
\affiliation{Physikalisch-Technische Bundesanstalt (PTB), 
Abbestr. 2-12, 10587 Berlin, Germany}

\date{\today}

\begin{abstract}
The geometry of a Si$_3$N$_4$ lamellar grating was investigated experimentally with reference-free grazing-incidence X-ray fluorescence analysis. While simple layered systems are usually treated with the matrix formalism to determine the X-ray standing wave field, this approach fails for laterally structured surfaces. Maxwell solvers based on finite elements are often used to model electrical field strengths for any 2D or 3D structures in the optical spectral range. We show that this approach can also be applied in the field of X-rays. The electrical field distribution obtained with the Maxwell solver can subsequently be used to calculate the fluorescence intensities in full analogy to the X-ray standing wave field obtained by the matrix formalism. Only the effective 1D integration for the layer system has to be replaced by a 2D integration of the finite elements, taking into account the local excitation conditions. We will show that this approach is capable of reconstructing the geometric line shape of a structured surface with high elemental sensitivity. This combination of GIXRF and finite-element simulations paves the way for a versatile characterization of nanoscale-structured surfaces.
\end{abstract}

\maketitle

\section{Introduction}

Recent progresses in different, related fields of science and industry relies on a well-controlled decrease in dimensions during the fabrication of various nanoscaled structures. In the semiconductor industry, this results in the integration of increasingly complex 2D and 3D architectures with feature sizes in the low-nm range\cite{S.Natarajan2014,markov_limits_2014}. 
A variety of other applications also benefit directly from the technological progress in the semiconductor industry. Surface-enhanced Raman scattering (SERS) \cite{A.X.Wang2015}, surface-enhanced infrared absorption spectroscopy (SEIRA) \cite{F.Le2008} and advanced light-trapping applications in solar cells \cite{Pala2013,M.L.Brongersma2014} are merely a few examples. Besides periodic lines as the simplest structure models, arrays of periodic nanostructures such as field emitter arrays \cite{Fletcher_2015} and nanorods \cite{Malerba2015} are also of great interest. 

The performance of such complex nm-structured devices depends on how well controlled both the dimensional parameters and the 3D elemental compositions are controlled within the nm structures themselves, as well as across large areas of structured substrate. Therefore, two different types of metrological tools that are capable of characterizing these measurands in a non-destructive way are required. These tools must have a high spatial resolution in order to provide insight on the distribution of the atoms within single nm-sized objects or be capable of probing large areas with sufficient sensitivity to the structure geometry that their average dimensional and chemical properties can be characterized. Emerging analytical techniques such as atom probe tomography (APT) provide sufficient lateral and chemical resolution to image single structures \cite{D.J.Larson2016}, but a non-destructive characterization of large numbers of objects by means of APT in order to gain statistical information is not possible. Grazing-incidence X-ray methods such as small-angle X-ray scattering (GISAXS) \cite{levine_grazing-incidence_1989} or grazing-incidence X-ray fluorescence (GIXRF) \cite{D.K.G.DeBoer1995,M.Dialameh2017,JAAS_2012}, are non-destructive ensemble methods and can easily probe larger sample areas with sufficient sensitivity to the dimensional and analytical parameters of the structures \cite{PhysRevLett.94.145504,hofmann_grazing_2009,rueda_grazing-incidence_2012,wernecke_direct_2012-1,gollmer_fabrication_2014,V.Soltwisch2016}. As X-ray fluorescence radiation is element-specific, GIXRF can provide compositional information in addition to scattering methods. When combined with a variation of the incident photon energy, even information on chemical binding sates can be gained by means of fluorescence detected X-ray absorption spectroscopy \cite{BEA_2015}.

In GIXRF, the incident angle $\theta_i$ between the X-ray beam and sample surface is typically varied around the critical angle $\theta_c$ for total external reflection. On flat samples, the interference between the incoming beam and the reflected beam results in an X-ray standing wave (XSW) field \cite{Bedzyk_1989,Golovchenko1982}, which can strongly modulate the intensity distribution above and below the reflecting surface depending on the specific layer structure. The intensity modulation inside the XSW field is correlated with the incident angle and the wavelength and significantly impacts the X-ray fluorescence intensity of an atom depending on its position within the XSW. Performing GIXRF angular scans thus provides information about the in-depth distribution of any probed element within the sample \cite{P.Hoenicke2009}. Employing radiometrically calibrated instrumentation \cite{Beckhoff2008} for reference-free GIXRF \cite{M.Mueller2014}, also provides quantitative information on the elemental mass deposition without the need for any external reference.

To model GIXRF angular profiles, an accurate calculation of the XSW intensity is essential. For a 1D system (for example a stratified layer stack) the recursive matrix formalism developed by Parratt \cite{Parratt1954} is often used and implemented in various software packages such as IMD \cite{Windt1998} and XSWini \cite{Pollakowski_2015}. This formalism is rather fast and is an ideal candidate for layered systems. However, if 2D or even complex 3D structures are present, these software packages are no longer capable of calculating the XSW. For special cases, e.g. for nanostructures with stochastic distribution on a surface, the approach of a stratified layer with reduced density has proven \cite{F.Reinhardt2014, M.Dialameh2017} to be sufficiently precise. Thus for a GIXRF-based characterization of regularly ordered 2D or 3D nanostructures, which are more relevant to fields such as the semiconductor industry, a novel calculation scheme is required for the XSW field, or in general for the near-field distribution.

Maxwell solvers based on the finite-element method (FEM) are suited for the computation of the electric near-field distribution (or in GIXRF terminology, the local excitation condition) within periodic arrangements of surface structures. They can thus contribute to the simulation and interpretation of GIXRF measurement data of such structures in order to derive the dimensional parameters of the structures as well as information about their elemental composition. Similar studies in the optical spectral range have demonstrated the potential of the {finite-element} method \cite{barth_2017}. Expanding this approach to include the X-ray spectral range is challenging due to the fact that the finite-element discretization of the computational domain necessary for this approach depends on the wave length of the incoming plane wave in order to ensure the numerical precision of the calculated electric-field distribution. For incident radiation with wavelengths in the nm or sub-nm range and domain sizes of several 100 nm, this seems to be only possible with a high computational effort, at first glance. But the orientation of the wave vector with respect to the geometrical layout of the sample defines the accessible numerical precision within a reasonable computation time. A more detailed analysis \cite{Soltwisch2017}, however showed that for the special orientation of the incoming wave vector in the GISAXS geometry these requirements relax and the computation becomes feasible.  

In this work, we demonstrate the flexibility and potential of the finite-element approach for the characterization of periodic structures using GIXRF and visualize the limitations of the Matrix method and the effective-layer approach. Experimental GIXRF results from a lithographically structured silicon nitride Si$_3$N$_4$ lamellar grating on a silicon substrate were compared to the first reconstruction results obtained with the Maxwell solver based on finite elements. The very good agreement between the measurements and the simulations, and the high sensitivity to relatively small changes in the geometrical layout, indicate the potential of the GIXRF method for the combined analytical and dimensional characterization of such nanostructured surfaces.

\section{Experimental Details}
\begin{figure}[htbp]
\centering
\includegraphics[width=0.48\textwidth]{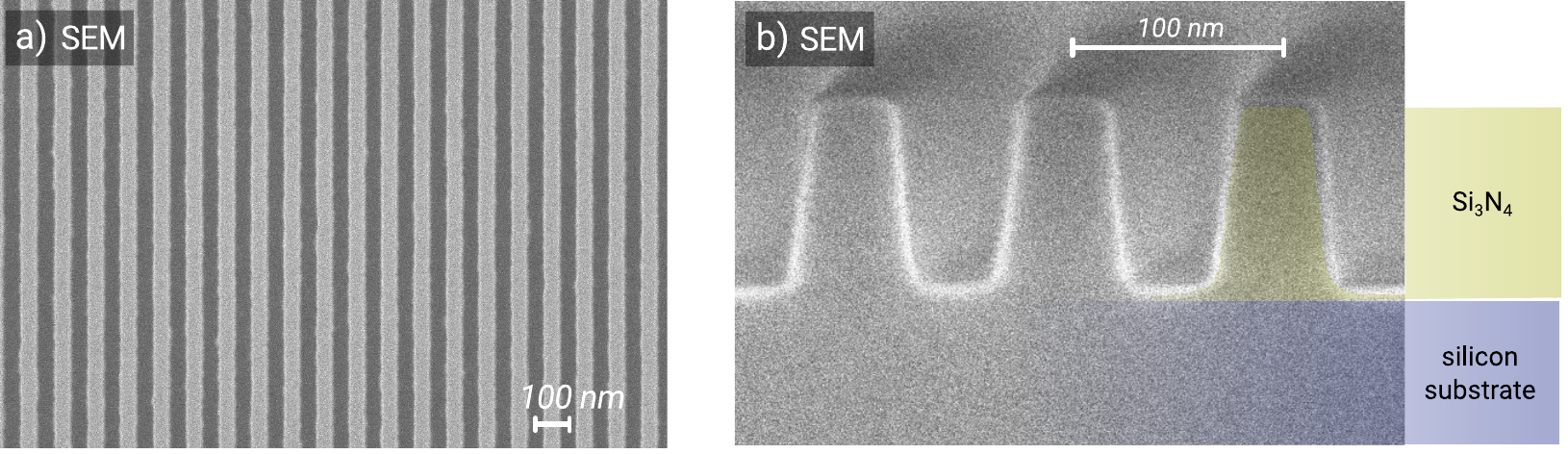}
	\caption{a) Scanning electron microscopy images (SEM) of the Si$_3$N$_4$ lamellar grating. b) SEM cross-section image obtained from a witness sample.}
\label{fig:sample}
\end{figure}

In this work, a lithographically structured silicon nitride layer on a silicon substrate was used. A Si$_3$N$_4$ lamellar grating was manufactured by means of electron beam lithography at the Helmholtz-Zentrum Berlin. The grating has a nominal pitch of 100 nm, a line height of 90 nm and a line width of 40 nm. The grating areas measure 1 mm by 15 mm, with the grating lines oriented parallel to the long edge. To manufacture the gratings, a silicon substrate with a 90 nm-thick Si$_3$N$_4$ layer was spin coated using ZEP520A the positive resist (organic polymer). The pattern was generated using a Vistec EBPG5000+ e-beam writer, operated with an electron acceleration voltage of 100 kV. After the resist development, the grating was etched into the Si$_3$N$_4$ layer via reactive ion etching using CHF$_3$. Finally, the remaining resist was removed by means of an oxygen plasma treatment. Scanning electron microscopy (SEM) images obtained from witness samples show the high quality of the periodic structured surface (see Fig.~\ref{fig:sample}).  

The reference-free GIXRF measurements were carried out at the plane-grating monochromator (PGM) beamline \cite{F.Senf1998} for undulator radiation in the PTB laboratory \cite{B.Beckhoff2009c} of the BESSY II synchrotron radiation facility. This beamline provides soft X-ray radiation of high spectral purity in the photon energy range of 78 eV to 1860 eV. The GIXRF experiments were conducted employing the radiometrically calibrated instrumentation \cite{Beckhoff2008} of PTB and a fundamental parameter-based reference-free quantification approach \cite{M.Mueller2014}. An ultrahigh-vacuum (UHV) chamber \cite{J.Lubeck2013} equipped with a 9-axis manipulator was used for the measurements, allowing the samples to be aligned precisely with respect to all necessary degrees of freedom. Photodiodes mounted on a 2$\theta$ axis allowed simultaneous X-ray reflectometry measurements to be performed and the samples to be aligned with respect to the incoming beam. The incident angle $\theta_i$ between the X-ray beam and the sample surface can be aligned with an uncertainty well below 0.01$^{\circ}$, which is sufficient for the GIXRF experiments. For structured surfaces, the azimuthal incidence angle $\varphi_i$ must also be taken into account. This is the angle between the lines of the grating structure and the plane of incidence defined by the incident beam and the normal to the sample surface, where $\varphi_i = 0^{\circ}$ corresponds to the position where the plane of incidence is parallel to the grating lines (conical). Under the prerequisite that at this position the grating structures be symmetrically with respect to the scattering plane, $\varphi_i$ can be directly aligned with the GIXRF signal to fit the conical mounting direction. The fluorescence radiation emitted is detected using a silicon drift detector (SDD) that has been calibrated with respect to its detector response functions and detection efficiency \cite{F.Scholze2009}. The incident photon flux is monitored by means of calibrated photodiodes.

The reference-free GIXRF experiments on the nanostructured Si$_3$N$_4$ lamellar gratings were performed using an incident photon energy of 520 eV. This photon energy is sufficiently high to excite the N-K${\alpha}$ fluorescence radiation, which serves here as a measure of the XSW intensity within the surface structure. It is also low enough to prevent the excitation of any O-K${\alpha}$ fluorescence radiation, which may complicate the spectral deconvolution of the fluorescence spectra. In addition to an angular variation of the incident angle $\theta_i$, the azimuthal angle $\varphi_i$ was also varied. The X-ray fluorescence spectra recorded for each $\theta_i$ and $\varphi_i$ combination were deconvoluted using detector response functions \cite{F.Scholze2009} for the relevant fluorescence lines.

The atomic fundamental parameters that quantitatively describe the process of absorption of the incident photons and the emission of the fluorescence photons are used together with the known instrumental parameters (e.g., the incident photon flux and the solid angle of detection) in order to also quantify the elemental mass deposition. In this work, we used an experimentally determined value for the nitrogen K-shell fluorescence yield in order to reduce the quantification uncertainty. This value was determined following the procedure described in ref.~\cite{P.Hoenicke2016a}.

\section{Simulation of fluorescence intensities}
\label{sec:fem}
\subsection{Maxwell solver based on the finite-element method}\label{sec_model}
\begin{figure}[tbp]
\centering
\includegraphics[width=0.38\textwidth]{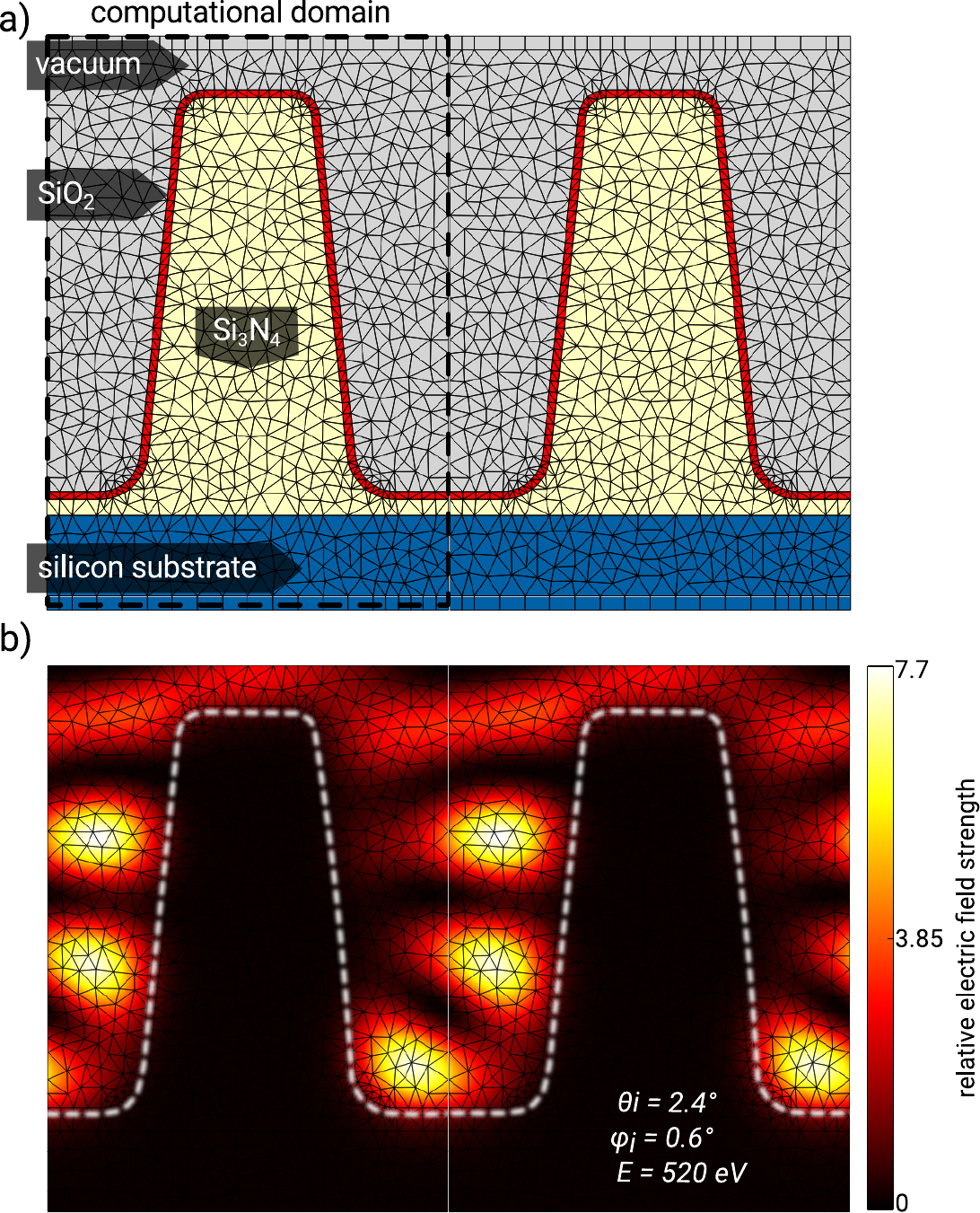}
	\caption{a) The computational domain of a lamellar grating model demonstrating the flexibility of the adaptive finite-element meshing algorithm. b) The computation of the electric field strength in false color scale, visualizing the formation of waveguide modes inside the grating grooves. The sample surface is indicated by a white dotted line.}
\label{fig:fem_mesh}
\end{figure}
X-rays are treated as electromagnetic plane waves of wavelength $\lambda= h c_0 / E_i$ (where $h=$ Planck's constant, $c_0=$ the speed of light and $E_i=$ the photon energy), which scatter on nanostructures. The set of Maxwell's equations can be rewritten as a single, second-order curl-curl equation for the electric field\cite{pomplun_adaptive_2007}. The general idea of the finite element discretization is that the computational domain is subdivided into small patches such as triangles. On these patches, a vectorial ansatz function is usually defined by means of polynomials that have a fixed order. The approximate electric field solution is the superposition of these local ansatz functions. Several software implementations of a Maxwell solver based on the finite-element approach are available. In this study, we use the JCMsuite package~\cite{pomplun_adaptive_2007}, which implements a higher-order finite-element method.

In Fig.~\ref{fig:fem_mesh}, a computational domain for the lamellar grating model (a) and the corresponding near-field simulation (b) used in this work are shown to demonstrate the flexibility of the finite-element meshing algorithm. The electric field distribution in Fig.~\ref{fig:fem_mesh} b) was simulated for incidence angles $\theta_i = 2.4^{\circ}$ and $\varphi_i=0.6^{\circ}$ and for a photon energy of 520 eV. The electric field distribution shows a clear difference between the layer and structured surface models. The appearance of nodes and anti-nodes inside the grating grooves below the critical angle of bulk Si$_3$N$_4$ ($\theta_c\sim 4^{\circ}$) gives rise to a significant surface-shape sensitivity of the integral signal observed. The penetration of the nodes into the grating structure, which increases the measured fluorescence emission, can be adjusted by varying both angles of incidence $\theta_i$ and $\varphi_i$. The finite-element meshing algorithm allows the line shape profile to be varied to form any geometrical layout. We have choosen a model close to the scanning electron microscopy (SEM) cross-section images we obtained from witness samples (see Fig.~\ref{fig:sample} b)). The line height, line width and sidewall angle were parameterized to allow the grating line profile to be easily changed. The line width of the Si$_3$N$_4$ grating is defined at the half-height of the finite-element model in order to eliminate the correlation between the sidewall angle and the structure density. The thickness of potential residual Si$_3$N$_4$ in the grooves is also implemented. To account for possible oxidization of the Si$_3$N$_4$ surface, which is well known from the oxygen plasma cleaning \cite{Kennedy_1999} of the photo resist stripping process, we included an additional SiO$_2$ layer with a homogeneous layer thickness covering the line profile (see Fig. \ref{fig:fem_mesh} a), red area).

\subsection{Layer approach for highly periodic nanostructured surfaces}
The simulation of the electric field distribution for the lamellar grating structure (Fig.~\ref{fig:fem_mesh} b)) indicates that the electric field intensity inside the grating bars, which stimulates the N-K$\alpha$ fluorescence emission, is weak but also directly coupled with both incidence angles. In literature, the so-called effective layer approach \cite{F.Reinhardt2014, M.Dialameh2017} is often used to describe the behavior of a structured surface. The structured surface is assumed to be a stratified layer with a reduced density. This can then be calculated by means of the Matrix method, requiring less numerical effort. 

\begin{figure}[tb]
\centering
\includegraphics[width=0.4\textwidth]{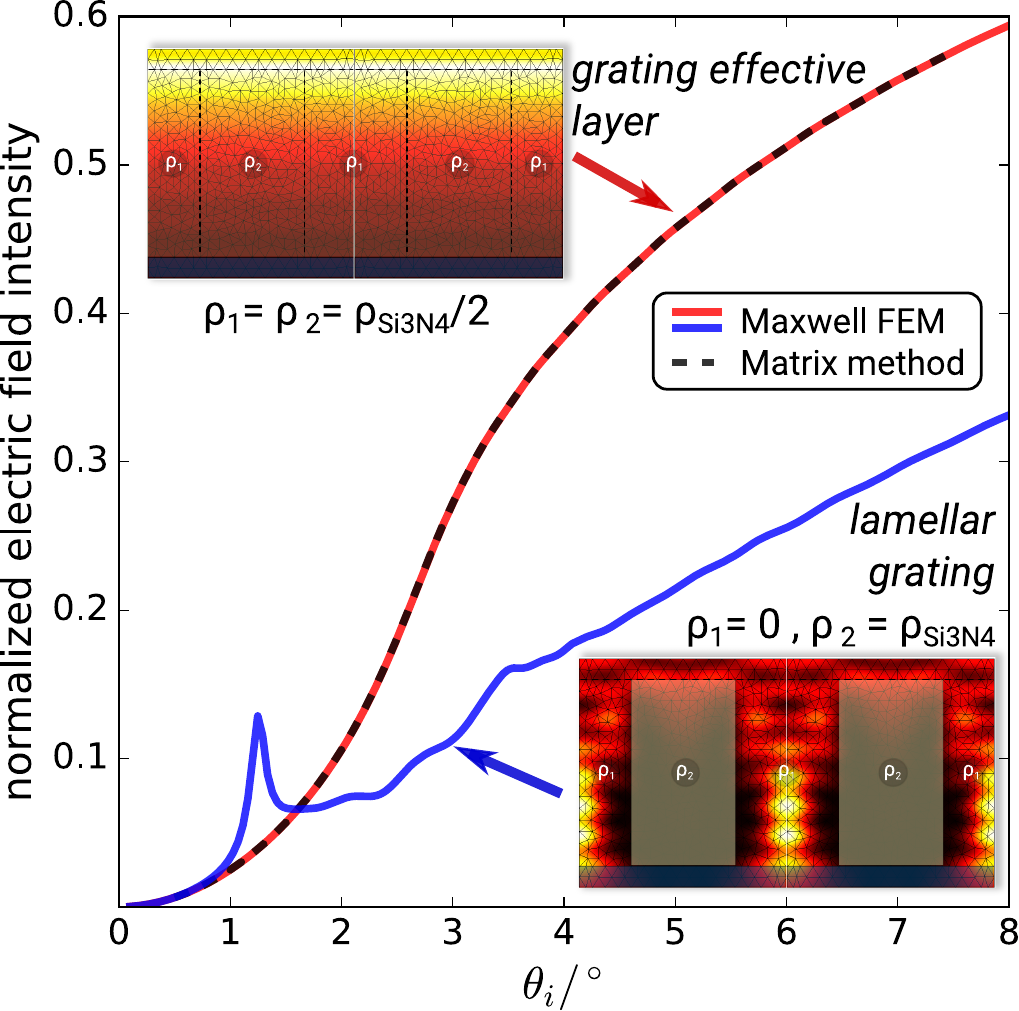}
\caption{Comparison between the simulation of electric field intensities in the Si$_3$N$_4$ domain from a grating effective layer system (red and black dashed line) with reduced density and the corresponding binary lamellar grating structure (blue line). The integrated field intensities were normalized to the incoming plane wave.}
\label{fig:XSW}
\end{figure}

However, the question arises as to whether a layer approach is capable of providing an approximate description of the fluorescence emission measured. In the first step, we thus compare the simulated electric field distribution of a grating effective layer system obtained by means of the Matrix method with that obtained by means of the Maxwell solver. This comparison is shown in Fig.~\ref{fig:XSW} (red and black dashed line) as a function of the grazing incidence angle $\theta_i$. A Si$_3$N$_4$ layer with a 90 nm thickness and half of the Si$_3$N$_4$ density ($\rho_1 = \rho_2 = \rho_{Si_{3}N_{4}}/2$) on top of a silicon substrate represents a grating effective layer of a perfect binary lamellar grating ($\rho_1 = 0, \rho_2 = \rho_{Si_{3}N_{4}}$). As expected, the results of the Maxwell solver computation are in perfect agreement with the results of the Matrix method for 1D layer systems. Both methods are able to deliver an adequate description of the fluorescence intensities measured for the layered systems. However, for the description of highly periodic structured surfaces with a simple layer approach, the effective layer approach and the Matrix method will fail. This is shown in the next step, where we simulate the field intensities of a binary lamellar grating structure (blue line). In contrast to the simulated angular electric field intensity profile of a Si$_3$N$_4$ layer with reduced density, which shows the expected kink only at the critical angle $\theta_c$ of Si$_3$N$_4$, the computation of the binary grating reveals a more complex intensity distribution, including several distinct features. 

The difference becomes obvious when the electric field distributions are compared directly for both approaches (see insets of Fig. \ref{fig:XSW}). They show the electric field distribution at an incident angle $\theta_i = 5^\circ$, where large differences can be observed. In the effective layer, the electric field distribution homogeneously penetrates the Si$_3$N$_4$, whereas the electric field is mainly confined to the space between the grating lines in the lamellar grating. From these simulations, it is clear that, for highly correlated systems, the effective layer approach is not able to describe the fluorescence emission measured. 

\subsection{Conversion of simulated electric field distributions into fluorescence intensities}
The deconvolved fluorescence intensities can be used to quantify the amount of material using the Sherman equation \cite{Sherman1955} in an adopted form for GIXRF \cite{P.Hoenicke2009}. The following equation applies to the fluorescence intensity $F(\theta_i,E_i)$ measured for 1D layer systems:
\begin{widetext}
\begin{equation}
F(\theta_i,E_i) = \frac{\Omega}{4\pi\sin\theta_i}\epsilon_{E_f} W_i \rho \omega_k\tau_{E_i} N_0 \int_0^{t_{max}} I(t,\theta_i,E_i) \cdot \exp\left[-t\rho\mu_{E_i}\right] dt\textrm{.}
\label{eq:sherman}
\end{equation}
\end{widetext}
The factor before the integral consists of fundamental, experimental and instrumental parameters. As all instrumental parameters are known due to our calibrated instrumentation \cite{Beckhoff2008,J.Lubeck2013}, we can use the calculated electric near-field intensity distribution or the XSW intensity distribution in GIXRF terminology $I(t,\theta_i,E_i)$ inside the grating structure to extract a numerical approximation of the expected fluorescence intensity $F(\theta_i,E_i)$ per incident photon. For this purpose, we interpolate the square of the absolute values of the computed electric field $|E(x,y)|^2$ distribution inside a specific area to a Cartesian grid $(x,y)$ with sufficient discretization ($dx\times dy=1$ nm$^2$). To account for self-attenuation, every field intensity on this grid is damped with respect to the path length $y_{dis}=(y-y_0) / \cos(\theta_i)$ of the emitted fluorescence photons through the Si$_3$N$_4$ in the direction of the fluorescence detector. Both the solid angle $\Omega/4\pi$ and the detection efficiency for N-K$\alpha$ radiation $\epsilon_{E_f}$ are known for this detector. The density $\rho$  and the attenuation coefficient $\mu_{E_i}$ for Si$_3$N$_4$ at the photon energy of the N-K$\alpha$ fluorescence line are taken from X-raylib\cite{T.Schoonjans2011} for bulk materials. 

We thus extend the Sherman equation to include 2D systems and numerically integrate $|E(x,y)|^2$  within the Si$_3$N$_4$ domain:
\begin{widetext}
\begin{equation}
\frac{4\pi\sin\theta_i}{\Omega}\frac{F(\theta_i,E_i)}{N_0\epsilon_{E_f}} =\frac{W_i \rho \tau(E_i) \omega_k}{\sum dx} \cdot \sum_{x}\sum_{y}  |E(x,y)|^2 \cdot \exp\left[-\rho\mu_{E_i}y_{dis}\right]\textrm{.} 
\end{equation}
\end{widetext}
This reformulation makes it possible to calculate the total emitted N-K$\alpha$ fluorescence intensity from the numerically calculated electrical field distributions and to compare it to the normalized experimental data. The mass fraction $W_i$ of nitrogen in Si$_3$N$_4$, as well as the fundamental parameters $\tau_{E_i}$ as the photo ionization cross section of the N-K shell \cite{T.Schoonjans2011} and $\omega_k$ as the fluorescence yield, are also required. $\omega_k$ was determined experimentally in a manner similar to that described for the O-K shell fluorescence yield in \cite{P.Hoenicke2016a}.

\subsection{Numerical accuracy and discretization size in the X-ray spectral range}
\begin{figure}[htbp]
\centering
\includegraphics[width=0.4\textwidth]{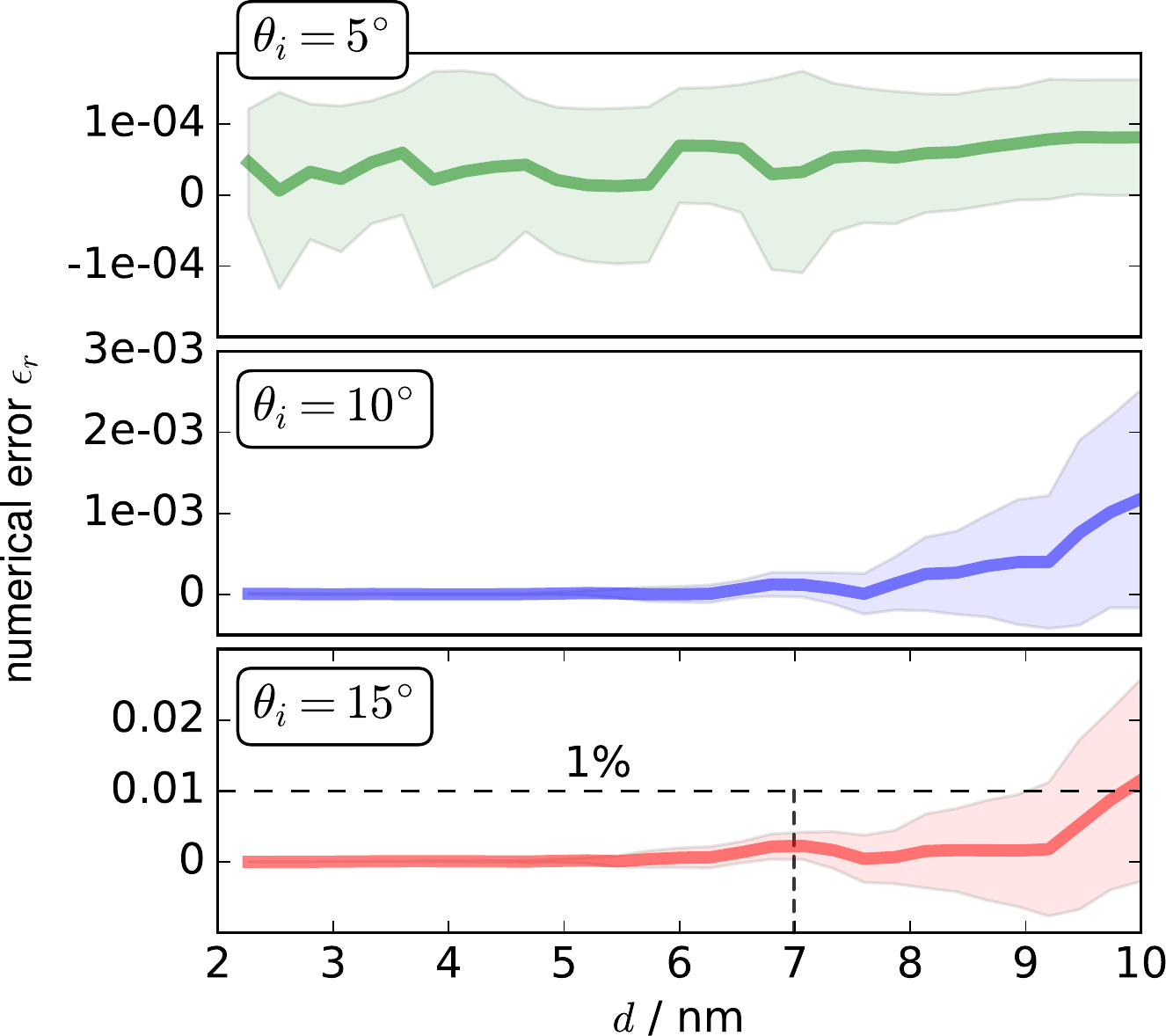}
	\caption{Convergence study of the numerical error ($E_i = 520$ eV, $p$=4) for the finite-element discretization length $d$ and three incidence angles $\theta_i$ ($5^{\circ},10^{\circ},15^{\circ}$). The shaded areas represent the 1$\sigma$ uncertainty due different geometrical models (see text). The horizontal dashed line indicates a relative numerical error of 1\%.}
\label{fig:convergence}
\end{figure}
The numerical accuracy of the approximate electric field is a function of the discretization size $d$ of the finite elements and the degree $p$ of the polynomials. The grazing incidence conical diffraction and the invariance of the grating in the direction of the scattering plane results in a standing wave field with much larger periodicity than the wavelength of the exciting radiation. This makes it possible to significantly increase the size of the required discretization length $d$ while still being in line with the conventional rule of half the relevant wavelength for the discretization to ensure numerical accuracy. 

In our experiment, the incident wavelength with $\lambda\approx 2.38$ nm (520 eV) combined with the conical scattering geometry and grazing incidence angles $\theta_i$ makes it possible to significantly increase the side lengths $d$ of the finite elements in the simulations. This enables the efficient use of a Maxwell solver based on the finite-element method to treat grazing incidence X-ray applications. Simulations based on a higher excitation energy for the investigation of other materials are also applicable.

This numerical stability is demonstrated in Fig.~\ref{fig:convergence}, which shows a convergence study of the numerical error as a function of the discretization length $d$ (with fixed polynomial order $p=4$) and of the incidence angle $\theta_i$. The absolute value of the relative numerical error $\epsilon_r$ is defined as
\begin{equation}
\epsilon_r (d,p,E,\theta_i,\varphi_i) = \frac{\left|I^{quasi}-I^{model}\right|}{I^{quasi}}\textrm{,}
\end{equation}
and is based on the computations of the integral electric field intensities within the Si$_3$N$_4$ domain $I^{model}$. The quasi-exact calculation $I^{quasi}$ is defined as the computation with the highest achievable numerical precision settings where further tuning of the precision parameters does not influence the results and a numerical stable solution is achieved. By increasing the numerical precision parameters, the $I^{model}$ calculation should converge against the quasi-exact solution. However, the incidence angles $\theta_i$ and $\varphi_i$ have a large impact on the numerical accuracy. At grazing incidence angles $\theta_i=5^{\circ}$ (see Fig.~\ref{fig:convergence}, green line), the numerical error $\epsilon_r$ is well below 10$^{-4}$, even with discretization lengths $d$ up to 10 nm, which is approximately four times the incidence wavelength. With increasing incidence angles $\theta_i=10^{\circ}$ (blue line) and $\theta_i=15^{\circ}$ (red line) in Fig.~\ref{fig:convergence}, the discretization length $d$ must be reduced to ensure a similar numerical precision. This is due to the fact that the spatial modulation frequency increases with increasing incidence angle $\theta_i$\cite{D.K.G.DeBoer1995}.

To account for changes in the numerical convergence due to different geometrical models, we simulated 1000 gratings by means of randomly distributed line shapes. The line height and line width was varied $\pm 10$ nm around the expected nominal values. The expected numerical error distribution \cite{Soltwisch2017} is shown in Fig.~\ref{fig:convergence} as shaded areas that represent the 1$\sigma$ interval. In the following reconstruction of the Si$_3$N$_4$ grating, we set the discretization length $d$ to 7 nm and the polynomial degree to 4, which is sufficient ($\epsilon_r < 1\%$) for the incidence angular range investigated in the GIXRF measurements. A dynamic incident angle-dependent adjustment of the discretization parameters during the simulation of GIXRF angular scans is possible, in principle, in order to further reduce the calculation time. However, for a simple 2D finite-element mesh with a 100 nm$^2$ domain, the computational time is well below 1 s for a single solution and sufficient for these first reconstruction attempts. Moreover, a constant remeshing of the identical structure is also computationally intensive.

\section{Reconstructing a Si$_3$N$_4$ lamellar grating by means of GIXRF}
\begin{figure}[htb]
\centering
\includegraphics[width=0.42\textwidth]{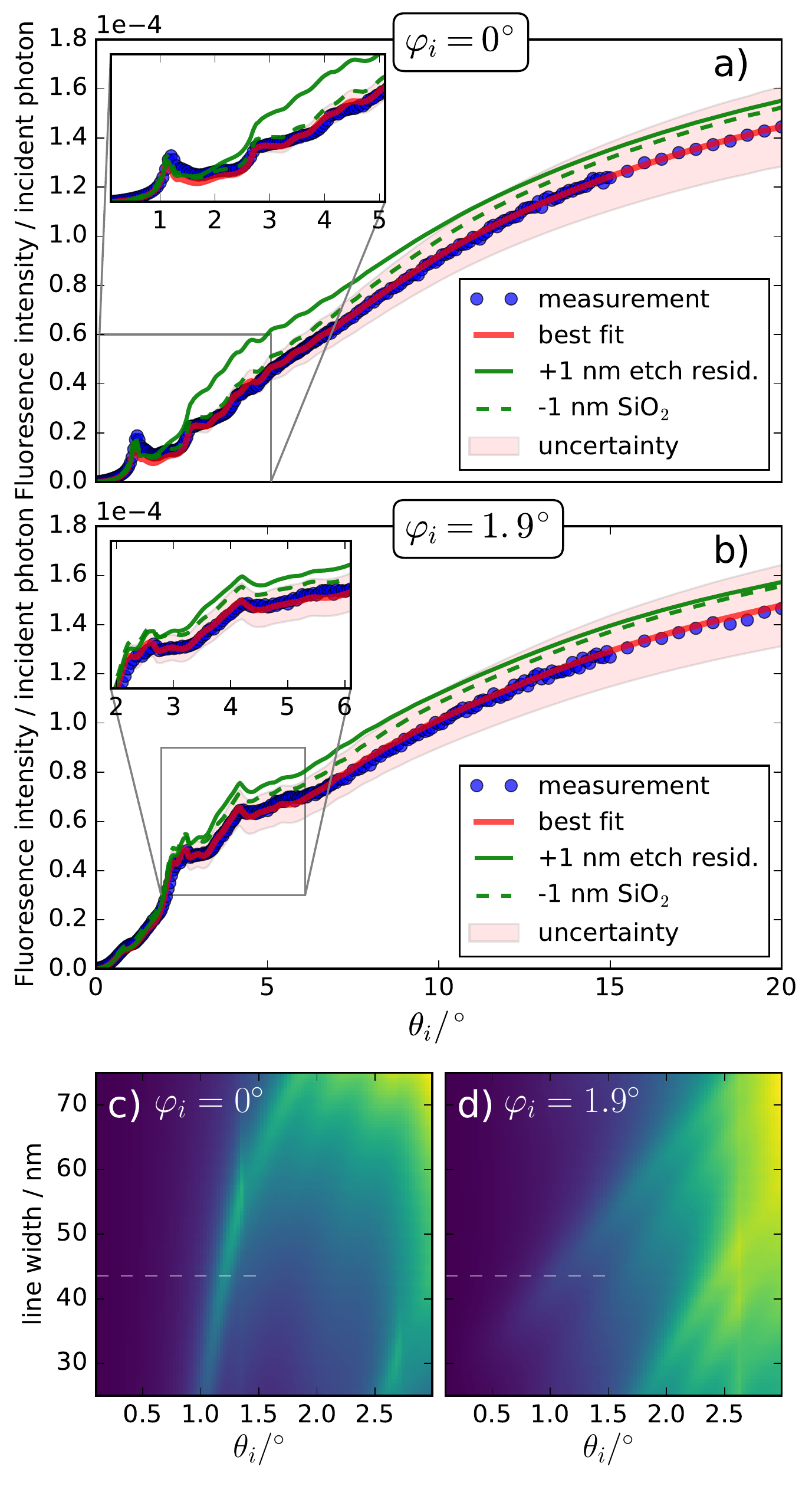}
	\caption{(a,b) Comparison between measured N-K$\alpha$ fluorescence intensity and the best reconstruction result for (a) conical mounting  $\varphi=0^{\circ}$ and (b) $\varphi=1.9^{\circ}$. Below the critical angle $\theta_c$ of bulk Si$_3$N$_4$, the intensity distribution is strongly modulated (inset in a) and b)). (c,d) Simulation of the $\theta_i$-dependent fluorescence intensity for line widths from 35 nm to 75 nm.}
\label{fig:lineplots_grating}
\end{figure}

The N-K$\alpha$ fluorescence intensity of the lamellar grating structure was determined for different combinations of $\theta_i$ and $\varphi_i$. Fig.~\ref{fig:lineplots_grating} a,b) shows the measured fluorescence intensity per incident photon with increasing incidence angle $\theta_i$ (blue dots) for two different azimuthal orientations $\varphi_i$.
At grazing incidence angles ($\theta_i < 5^{\circ}$), the fluorescence intensity emitted exhibits a rather complex modulation (see insets in Fig.~\ref{fig:lineplots_grating}) that can no longer be explained with the grating effective layer model (see also Fig.~\ref{fig:XSW}). The structure-related impact on the intensity distribution is also highly visible if the azimuthal incident angle $\varphi_i$ is shifted from the perfect conical mounting of the grating (cf.~Fig.~\ref{fig:lineplots_grating}a and Fig.~\ref{fig:lineplots_grating}b). Both curves show a completely different intensity modulation at grazing incidence angles below the critical angle $\theta_c$ of total external reflection of the Si$_3$N$_4$ grating. The electric near-field calculation of the grating model (see Fig.~\ref{fig:fem_mesh}b) reveal that the electric field is strongly located inside the grooves until the critical angle $\theta_c$ is reached and the field begins to penetrate the grating bars. This feature provides a high sensitivity to the surface shape for periodically structured surfaces in GIXRF experiments. 

In this study, we are not yet able to give a full reconstruction of the line shape model including uncertainties based on the Maxwell solver and the finite-element approach. A statistical evaluation of all model parameters (see Section \ref{sec_model}) is possible, in principle but requires further optimization of the simulation algorithm to reduce the computational effort. In Fig.~\ref{fig:lineplots_grating} a) and b), the red lines represent the best fit obtained with a Monte Carlo method. The starting values for the expected grating model parameters are based on the SEM cross-sections measurements from the witness samples. The 1$\sigma$ uncertainty bands (red shaded areas) in Fig.~\ref{fig:lineplots_grating} are calculated only using the uncertainties from the experimentally determined fluorescence yield $\omega_k$ and the photo ionization cross section $\tau_{E_i}$, as their contributions are expected to be dominant. Their combined uncertainty is about 11 \%, which makes it possible to neglect any other experimental or numerical uncertainty.

Even though we have not performed a full reconstruction of the line shape model including uncertainties, several conclusions can be drawn from the optimization we performed. First, the grating structure is fully etched down to the substrate, thus making the thickness of the Si$_3$N$_4$ in the grooves zero. The method is very sensitive to this parameter, as even a very thin remaining layer imposes large changes on the calculated signal. In Fig.~\ref{fig:lineplots_grating} a,b) the green solid line shows the fluorescence intensity obtained with the best model including an additional 1 nm thick etch residual. Second, an oxidized layer on the surface of the  Si$_3$N$_4$ is definitely present. From the optimization performed, we cannot conclude whether it is a pure SiO$_2$ layer or something else that results in a depletion of nitrogen at the surface; However, an SiO$_2$ layer is very likely to exist due to the oxygen plasma cleaning \cite{Kennedy_1999}. The optimized thickness of this layer is 3.5 nm, assuming a box-like depth profile. A gradient profile that is more realistic could be included as well, although this would increase the numerical effort and was thus not taken into account here. However, the sensitivity of the method for this parameter is also very high (see Fig.~\ref{fig:lineplots_grating} a,b) the green dashed line where the thickness of the oxygen layer has been reduced by 1 nm). By additionally measuring the fluorescence emission from oxygen, this could even be improved in the future.


Due to the quantitative modeling enabled by the reference-free GIXRF, we can derive the mass deposition and thus the density of the Si$_3$N$_4$ layer. From the optimization performed, it is determined to be approximately $\rho = 2.8 \frac{g}{cm^2}$, which is 10 \% below literature values from Si$_3$N$_4$ thin films. It should be noted that this is influenced by the simplified model of a well-separated oxidization layer, as most of the nitrogen fluorescence signal measured is generated in close proximity to the surface of the structure (see Fig.~\ref{fig:fem_mesh}b).

The other reconstruction results (line height at $\sim 87$ nm, line width at $\sim 43$ nm and sidewall angle at $\sim 86^{\circ}$) are in good agreement with the expected nominal parameters and the SEM cross-section images obtained from witness samples (see Fig.~\ref{fig:sample}). For these dimensional parameters of the grating, the GIXRF methodology presented is also provides a good sensitivity. This is demonstrated in Fig.~\ref{fig:lineplots_grating} c) and d), where the calculated nitrogen fluorescence angular profile is shown in a false color scale as a function of the line width of the grating for two different azimuthal angles $\varphi_i$. In both geometries, the intensity distribution and the position of the different features in the GIXRF angular distributions is highly correlated with the line shape of the lamellar grating structure.\\

\begin{figure*}[htb]
\centering
\includegraphics[width=0.85\textwidth]{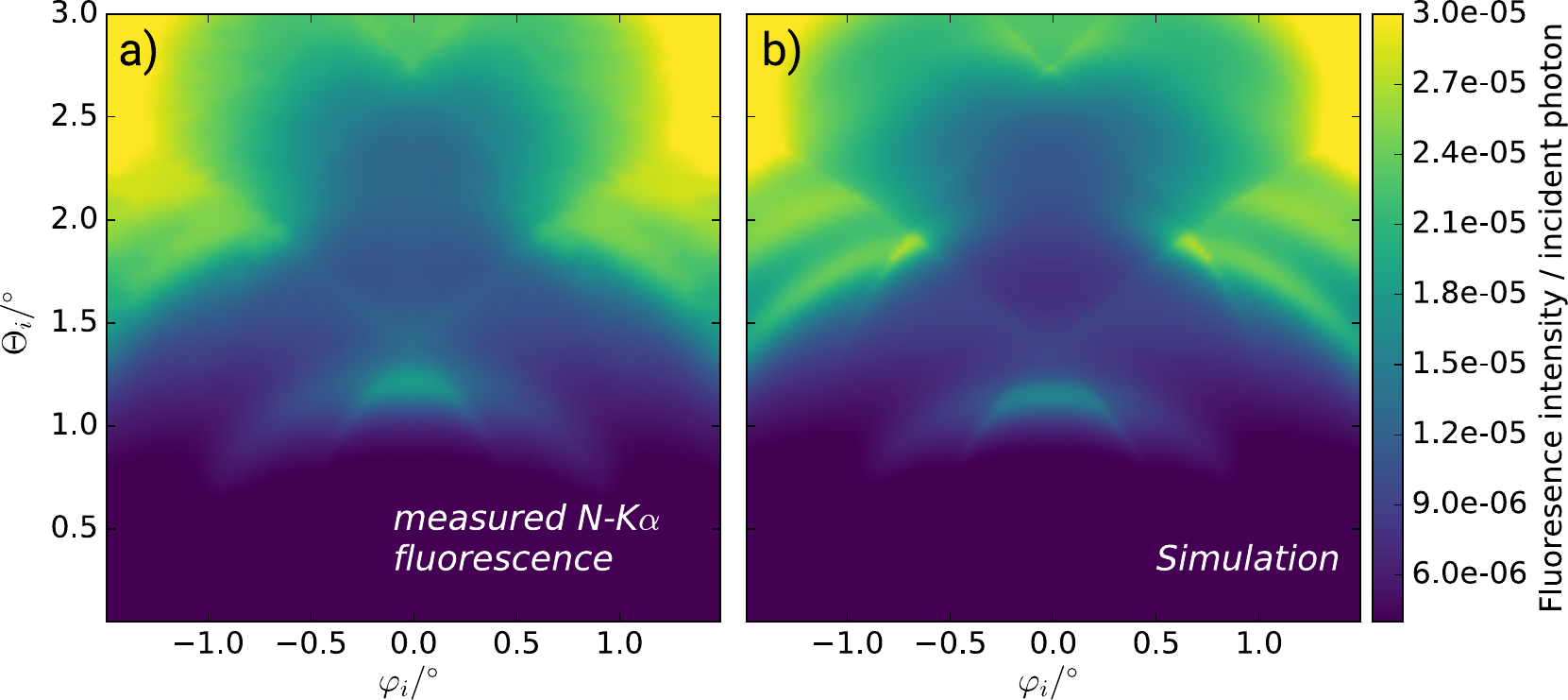}
	\caption{Comparison between the measured fluorescence intensity map (a) of the Si$_3$N$_4$ grating under various incidence angles ($\theta_i,\varphi_i$) and the simulated fluorescence map (b) based on a line shape model with the best reconstruction result.}
\label{fig:cdsen}
\end{figure*}

In Fig.~\ref{fig:cdsen}, the full angular measurement dataset (a) is shown in comparison to the simulation (b). Here, the incident angles were varied from $\theta_i = 0^{\circ}$ to $3^{\circ}$ (y-axis) and the azimuthal angle was varied between $\varphi_i = -0.1^{\circ}$ and $1.5^{\circ}$ (x-axis). Due to the symmetry at $\varphi_i = 0^{\circ}$, the experimental data is also valid for negative $\varphi_i$ angles. In Fig.~\ref{fig:cdsen} b), the simulation using the best reconstruction result obtained with the Maxwell solver is shown for the same angular ranges. Several distinct features are visible in both fluorescence maps. The formation of perfect circles around the symmetry axis corresponds to the penetration of the XSW nodes inside the Si$_3$N$_4$ structure. The overall agreement between experimental data and the calculation result is very good for the full angular ranges. The beam divergence was not included in the theoretical simulation in order not to degrade the fine details of the fluorescence map (for example the sharply curved lines that are very similar to higher-order Yoneda lines observed in GISAXS experiments \cite{V.Soltwisch2016}). These details are linked to the periodicity of the nanostructured surface.

\section{Conclusions}
In this work, we have shown that a GIXRF-based characterization of regularly ordered, nanostructured surfaces requires a finite element-based calculation scheme in order to model the experimental data. While simulations based on the conventional matrix formalism allow a GIXRF investigation of simple layered systems to take place, this approach will fail if a periodic surface structure such as a grating is present. In this case, the interference due to the periodic arrangement must be taken into account in the model. We have shown that Maxwell solvers based on finite elements are ideally suited for calculating the electric field intensities of any 2D or 3D structure. This allows a GIXRF-based characterization of regularly ordered nanoscale structured surfaces to take place, thus making it an interesting and novel approach with great potential in the different fields of nanotechnology. 

We have applied the reference-free GIXRF technique of PTB to a nanoscale lamellar grating consisting of Si$_3$N$_4$ on Si. A finite element-based simulation for both incident angle-dependent intensity distributions within the nanostructures is used to model the experimental GIXRF data, and the dimensional parameters of the grating as well as the elemental distributions are derived. Even though only a rough model of the experimental data has been provided here, we have shown that this technique provides a direct access to the spatial distribution with promising sensitivity for the characterization of these parameters. For the example presented, this sensitivity could even be enhanced by changing to or adding a higher excitation photon energy in order to also gain a fluorescence signal from the oxide layer on the surface. A further improvement of the numerical accuracy is possible by implementing the self-attenuation correction directly to the integration of the finite elements. In addition, the flexibility of the finite elements provides an opportunity to gain deeper insight into the elemental distribution of the nanostructures investigated. For example, it is possible to model 2D structures including complex interdiffusion layers. The GIXRF technique can also be combined with other techniques such as grazing-incidence small-angle X-ray scattering, X-ray reflectometry or resonant X-ray scattering in future experiments to take advantage of their complementary nature \cite{Haase2016}.

In summary, reference-free GIXRF is clearly suitable as a new, non-destructive metrology tool for the dimensional and elemental characterization of nanostructured surfaces. In principle, this technique is also transferable to laboratory scale tools for GIXRF if an appropriate calibration is available and if quantitative information can be derived from the fluorescence intensities measured. As such nanostructures are of rapidly increasing relevance in many fields of applications, the technique presented here is of great interest to the field of nanotechnology.

\section{Conflicts of interest}
There are no conflicts of interest to declare.

\section{Acknowledgements}
Parts of this research were performed within the ‘3D MetChemIT’ EMPIR project. The financial support of the EMPIR program is gratefully acknowledged. Our project is jointly funded by the European Metrology Programme for Innovation and Research (EMPIR) and participating countries within the European Association of National Metrology Institutes (EURAMET) and the European Union.

\bibliography{lit} 

\providecommand*{\mcitethebibliography}{\thebibliography}
\csname @ifundefined\endcsname{endmcitethebibliography}
{\let\endmcitethebibliography\endthebibliography}{}
\begin{mcitethebibliography}{41}
\providecommand*{\natexlab}[1]{#1}
\providecommand*{\mciteSetBstSublistMode}[1]{}
\providecommand*{\mciteSetBstMaxWidthForm}[2]{}
\providecommand*{\mciteBstWouldAddEndPuncttrue}
  {\def\EndOfBibitem{\unskip.}}
\providecommand*{\mciteBstWouldAddEndPunctfalse}
  {\let\EndOfBibitem\relax}
\providecommand*{\mciteSetBstMidEndSepPunct}[3]{}
\providecommand*{\mciteSetBstSublistLabelBeginEnd}[3]{}
\providecommand*{\EndOfBibitem}{}
\mciteSetBstSublistMode{f}
\mciteSetBstMaxWidthForm{subitem}
{(\emph{\alph{mcitesubitemcount}})}
\mciteSetBstSublistLabelBeginEnd{\mcitemaxwidthsubitemform\space}
{\relax}{\relax}

\bibitem[Natarajan \emph{et~al.}(2014)Natarajan, Agostinelli, Akbar, Bost,
  Bowonder, Chikarmane, Chouksey, Dasgupta, Fischer, Fu, Ghani, Giles,
  S.~Govindaraju, Han, Hanken, Haralson, Haran, Heckscher, Heussner, Jain,
  James, Jhaveri, Jin, Kam, Karl, Kenyon, Liu, Luo, Mehandru, Morarka, Neiberg,
  Packan, Paliwal, Parker, Patel, Patel, Pelto, Pipes, Plekhanov, Prince,
  Rajamani, Sandford, Sell, Sivakumar, Smith, Song, Tone, Troeger, Wiedemer,
  Yang, and Zhang]{S.Natarajan2014}
S.~Natarajan, M.~Agostinelli, S.~Akbar, M.~Bost, A.~Bowonder, V.~Chikarmane,
  S.~Chouksey, A.~Dasgupta, K.~Fischer, Q.~Fu, T.~Ghani, M.~Giles, R.~G.
  S.~Govindaraju, W.~Han, D.~Hanken, E.~Haralson, M.~Haran, M.~Heckscher,
  R.~Heussner, P.~Jain, R.~James, R.~Jhaveri, I.~Jin, H.~Kam, E.~Karl,
  C.~Kenyon, M.~Liu, Y.~Luo, R.~Mehandru, S.~Morarka, L.~Neiberg, P.~Packan,
  A.~Paliwal, C.~Parker, P.~Patel, R.~Patel, C.~Pelto, L.~Pipes, P.~Plekhanov,
  M.~Prince, S.~Rajamani, J.~Sandford, B.~Sell, S.~Sivakumar, P.~Smith,
  B.~Song, K.~Tone, T.~Troeger, J.~Wiedemer, M.~Yang and K.~Zhang,
  \emph{Electron Devices Meeting (IEDM), 2014 IEEE International}, 2014,  3.7.1
  -- 3.7.3\relax
\mciteBstWouldAddEndPuncttrue
\mciteSetBstMidEndSepPunct{\mcitedefaultmidpunct}
{\mcitedefaultendpunct}{\mcitedefaultseppunct}\relax
\EndOfBibitem
\bibitem[Markov(2014)]{markov_limits_2014}
I.~L. Markov, \emph{Nature}, 2014, \textbf{512}, 147--154\relax
\mciteBstWouldAddEndPuncttrue
\mciteSetBstMidEndSepPunct{\mcitedefaultmidpunct}
{\mcitedefaultendpunct}{\mcitedefaultseppunct}\relax
\EndOfBibitem
\bibitem[Wang and Kong(2015)]{A.X.Wang2015}
A.~Wang and X.~Kong, \emph{Materials}, 2015, \textbf{8}, 3024--3052\relax
\mciteBstWouldAddEndPuncttrue
\mciteSetBstMidEndSepPunct{\mcitedefaultmidpunct}
{\mcitedefaultendpunct}{\mcitedefaultseppunct}\relax
\EndOfBibitem
\bibitem[Le \emph{et~al.}(2008)Le, Brandl, Urzhumov, Wang, Kundu, Halas,
  Aizpurua, and Nordlander]{F.Le2008}
F.~Le, D.~Brandl, Y.~Urzhumov, H.~Wang, J.~Kundu, N.~Halas, J.~Aizpurua and
  P.~Nordlander, \emph{ACS Nano}, 2008, \textbf{2}, 707--708\relax
\mciteBstWouldAddEndPuncttrue
\mciteSetBstMidEndSepPunct{\mcitedefaultmidpunct}
{\mcitedefaultendpunct}{\mcitedefaultseppunct}\relax
\EndOfBibitem
\bibitem[Pala \emph{et~al.}(2013)Pala, Liu, Barnard, Askarov, Garnett, Fan, and
  Brongersma]{Pala2013}
R.~Pala, J.~Liu, E.~Barnard, D.~Askarov, E.~Garnett, S.~Fan and M.~Brongersma,
  \emph{Nat. Commun.}, 2013, \textbf{4}, 2095\relax
\mciteBstWouldAddEndPuncttrue
\mciteSetBstMidEndSepPunct{\mcitedefaultmidpunct}
{\mcitedefaultendpunct}{\mcitedefaultseppunct}\relax
\EndOfBibitem
\bibitem[Brongersma \emph{et~al.}(2014)Brongersma, Cui, and
  Fan]{M.L.Brongersma2014}
M.~Brongersma, Y.~Cui and S.~Fan, \emph{Nat Mater}, 2014, \textbf{13},
  451--460\relax
\mciteBstWouldAddEndPuncttrue
\mciteSetBstMidEndSepPunct{\mcitedefaultmidpunct}
{\mcitedefaultendpunct}{\mcitedefaultseppunct}\relax
\EndOfBibitem
\bibitem[Fletcher \emph{et~al.}(2013)Fletcher, Mangalam, Martin, and
  King]{Fletcher_2015}
P.~C. Fletcher, V.~K.~R. Mangalam, L.~W. Martin and W.~P. King, \emph{Journal
  of Vacuum Science \& Technology B, Nanotechnology and Microelectronics:
  Materials, Processing, Measurement, and Phenomena}, 2013, \textbf{31},
  021805\relax
\mciteBstWouldAddEndPuncttrue
\mciteSetBstMidEndSepPunct{\mcitedefaultmidpunct}
{\mcitedefaultendpunct}{\mcitedefaultseppunct}\relax
\EndOfBibitem
\bibitem[Malerba \emph{et~al.}(2015)Malerba, Alabastri, Miele, Zilio, Patrini,
  Bajoni, Messina, Dipalo, Toma, Proietti~Zaccaria, and
  De~Angelis]{Malerba2015}
M.~Malerba, A.~Alabastri, E.~Miele, P.~Zilio, M.~Patrini, D.~Bajoni, G.~C.
  Messina, M.~Dipalo, A.~Toma, R.~Proietti~Zaccaria and F.~De~Angelis, 2015,
  \textbf{5}, 16436 EP --\relax
\mciteBstWouldAddEndPuncttrue
\mciteSetBstMidEndSepPunct{\mcitedefaultmidpunct}
{\mcitedefaultendpunct}{\mcitedefaultseppunct}\relax
\EndOfBibitem
\bibitem[Larson \emph{et~al.}(2016)Larson, Prosa, Perea, Inoue, and
  Mangelinck]{D.J.Larson2016}
D.~Larson, T.~Prosa, D.~Perea, K.~Inoue and D.~Mangelinck, \emph{MRS Bulletin},
  2016, \textbf{41(1)}, 30--34\relax
\mciteBstWouldAddEndPuncttrue
\mciteSetBstMidEndSepPunct{\mcitedefaultmidpunct}
{\mcitedefaultendpunct}{\mcitedefaultseppunct}\relax
\EndOfBibitem
\bibitem[Levine \emph{et~al.}(1989)Levine, Cohen, Chung, and
  Georgopoulos]{levine_grazing-incidence_1989}
J.~R. Levine, J.~B. Cohen, Y.~W. Chung and P.~Georgopoulos, \emph{J. Appl.
  Cryst.}, 1989, \textbf{22}, 528--532\relax
\mciteBstWouldAddEndPuncttrue
\mciteSetBstMidEndSepPunct{\mcitedefaultmidpunct}
{\mcitedefaultendpunct}{\mcitedefaultseppunct}\relax
\EndOfBibitem
\bibitem[de~Boer \emph{et~al.}(1995)de~Boer, Leenaers, and van~den
  Hoogenhof]{D.K.G.DeBoer1995}
D.~de~Boer, A.~Leenaers and W.~van~den Hoogenhof, \emph{X-Ray Spectrom.}, 1995,
  \textbf{24(3)}, 91--102\relax
\mciteBstWouldAddEndPuncttrue
\mciteSetBstMidEndSepPunct{\mcitedefaultmidpunct}
{\mcitedefaultendpunct}{\mcitedefaultseppunct}\relax
\EndOfBibitem
\bibitem[Dialameh \emph{et~al.}(2017)Dialameh, Lupi, Hönicke, Kayser,
  Beckhoff, Weimann, Fleischmann, Vandervorst, Dubcek, Pivac, Perego, Seguini,
  Leo, and Boarino]{M.Dialameh2017}
M.~Dialameh, F.~F. Lupi, P.~Hönicke, Y.~Kayser, B.~Beckhoff, T.~Weimann,
  C.~Fleischmann, W.~Vandervorst, P.~Dubcek, B.~Pivac, M.~Perego, G.~Seguini,
  N.~D. Leo and L.~Boarino, \emph{Physica Status Solidi}, 2017\relax
\mciteBstWouldAddEndPuncttrue
\mciteSetBstMidEndSepPunct{\mcitedefaultmidpunct}
{\mcitedefaultendpunct}{\mcitedefaultseppunct}\relax
\EndOfBibitem
\bibitem[H\"onicke \emph{et~al.}(2012)H\"onicke, Kayser, Beckhoff, M\"uller,
  Dousse, Hoszowska, and Nowak]{JAAS_2012}
P.~H\"onicke, Y.~Kayser, B.~Beckhoff, M.~M\"uller, J.~Dousse, J.~Hoszowska and
  S.~Nowak, \emph{J. Anal. At. Spectrom.}, 2012, \textbf{27}, 1432--1438\relax
\mciteBstWouldAddEndPuncttrue
\mciteSetBstMidEndSepPunct{\mcitedefaultmidpunct}
{\mcitedefaultendpunct}{\mcitedefaultseppunct}\relax
\EndOfBibitem
\bibitem[Narayanan \emph{et~al.}(2005)Narayanan, Lee, Guico, Sinha, and
  Wang]{PhysRevLett.94.145504}
S.~Narayanan, D.~R. Lee, R.~S. Guico, S.~K. Sinha and J.~Wang, \emph{Phys. Rev.
  Lett.}, 2005, \textbf{94}, 145504\relax
\mciteBstWouldAddEndPuncttrue
\mciteSetBstMidEndSepPunct{\mcitedefaultmidpunct}
{\mcitedefaultendpunct}{\mcitedefaultseppunct}\relax
\EndOfBibitem
\bibitem[Hofmann \emph{et~al.}(2009)Hofmann, Dobisz, and
  Ocko]{hofmann_grazing_2009}
T.~Hofmann, E.~Dobisz and B.~M. Ocko, \emph{J. Vac. Sci. Technol. B}, 2009,
  \textbf{27}, 3238--3243\relax
\mciteBstWouldAddEndPuncttrue
\mciteSetBstMidEndSepPunct{\mcitedefaultmidpunct}
{\mcitedefaultendpunct}{\mcitedefaultseppunct}\relax
\EndOfBibitem
\bibitem[Rueda \emph{et~al.}(2012)Rueda, Martín-Fabiani, Soccio, Alayo,
  Pérez-Murano, Rebollar, García-Gutiérrez, Castillejo, and
  Ezquerra]{rueda_grazing-incidence_2012}
D.~R. Rueda, I.~Martín-Fabiani, M.~Soccio, N.~Alayo, F.~Pérez-Murano,
  E.~Rebollar, M.~C. García-Gutiérrez, M.~Castillejo and T.~A. Ezquerra,
  \emph{J. Appl. Cryst.}, 2012, \textbf{45}, 1038--1045\relax
\mciteBstWouldAddEndPuncttrue
\mciteSetBstMidEndSepPunct{\mcitedefaultmidpunct}
{\mcitedefaultendpunct}{\mcitedefaultseppunct}\relax
\EndOfBibitem
\bibitem[Wernecke \emph{et~al.}(2012)Wernecke, Scholze, and
  Krumrey]{wernecke_direct_2012-1}
J.~Wernecke, F.~Scholze and M.~Krumrey, \emph{Rev. Sci. Instrum.}, 2012,
  \textbf{83}, 103906\relax
\mciteBstWouldAddEndPuncttrue
\mciteSetBstMidEndSepPunct{\mcitedefaultmidpunct}
{\mcitedefaultendpunct}{\mcitedefaultseppunct}\relax
\EndOfBibitem
\bibitem[Gollmer \emph{et~al.}(2014)Gollmer, Walter, Lorch, Novák, Banerjee,
  Dieterle, Santoro, Schreiber, Kern, and Fleischer]{gollmer_fabrication_2014}
D.~A. Gollmer, F.~Walter, C.~Lorch, J.~Novák, R.~Banerjee, J.~Dieterle,
  G.~Santoro, F.~Schreiber, D.~P. Kern and M.~Fleischer, \emph{Micro. Engine.},
  2014, \textbf{119}, 122--126\relax
\mciteBstWouldAddEndPuncttrue
\mciteSetBstMidEndSepPunct{\mcitedefaultmidpunct}
{\mcitedefaultendpunct}{\mcitedefaultseppunct}\relax
\EndOfBibitem
\bibitem[Soltwisch \emph{et~al.}(2016)Soltwisch, Haase, Wernecke, Probst,
  Schoengen, Burger, Krumrey, and Scholze]{V.Soltwisch2016}
V.~Soltwisch, A.~Haase, J.~Wernecke, J.~Probst, M.~Schoengen, S.~Burger,
  M.~Krumrey and F.~Scholze, \emph{Phys. Rev. B}, 2016, \textbf{94},
  035419\relax
\mciteBstWouldAddEndPuncttrue
\mciteSetBstMidEndSepPunct{\mcitedefaultmidpunct}
{\mcitedefaultendpunct}{\mcitedefaultseppunct}\relax
\EndOfBibitem
\bibitem[Pollakowski and Beckhoff(2015)]{BEA_2015}
B.~Pollakowski and B.~Beckhoff, \emph{Anal. Chem.}, 2015, \textbf{87(15)},
  7705--7711\relax
\mciteBstWouldAddEndPuncttrue
\mciteSetBstMidEndSepPunct{\mcitedefaultmidpunct}
{\mcitedefaultendpunct}{\mcitedefaultseppunct}\relax
\EndOfBibitem
\bibitem[Bedzyk \emph{et~al.}(1989)Bedzyk, Bommarito, and
  Schildkraut]{Bedzyk_1989}
M.~J. Bedzyk, G.~M. Bommarito and J.~S. Schildkraut, \emph{Phys. Rev. Lett.},
  1989, \textbf{62}, 1376--1379\relax
\mciteBstWouldAddEndPuncttrue
\mciteSetBstMidEndSepPunct{\mcitedefaultmidpunct}
{\mcitedefaultendpunct}{\mcitedefaultseppunct}\relax
\EndOfBibitem
\bibitem[Golovchenko \emph{et~al.}(1982)Golovchenko, Patel, Kaplan, Cowan, and
  Bedzyk]{Golovchenko1982}
J.~A. Golovchenko, J.~R. Patel, D.~R. Kaplan, P.~L. Cowan and M.~J. Bedzyk,
  \emph{Phys. Rev. Lett.}, 1982, \textbf{49}, 560--563\relax
\mciteBstWouldAddEndPuncttrue
\mciteSetBstMidEndSepPunct{\mcitedefaultmidpunct}
{\mcitedefaultendpunct}{\mcitedefaultseppunct}\relax
\EndOfBibitem
\bibitem[Hönicke \emph{et~al.}(2010)Hönicke, Beckhoff, Kolbe, Giubertoni,
  van~den Berg, and Pepponi]{P.Hoenicke2009}
P.~Hönicke, B.~Beckhoff, M.~Kolbe, D.~Giubertoni, J.~van~den Berg and
  G.~Pepponi, \emph{Anal. Bioanal. Chem.}, 2010, \textbf{396(8)},
  2825--2832\relax
\mciteBstWouldAddEndPuncttrue
\mciteSetBstMidEndSepPunct{\mcitedefaultmidpunct}
{\mcitedefaultendpunct}{\mcitedefaultseppunct}\relax
\EndOfBibitem
\bibitem[Beckhoff(2008)]{Beckhoff2008}
B.~Beckhoff, \emph{J. Anal. At. Spectrom.}, 2008, \textbf{23}, 845 -- 853\relax
\mciteBstWouldAddEndPuncttrue
\mciteSetBstMidEndSepPunct{\mcitedefaultmidpunct}
{\mcitedefaultendpunct}{\mcitedefaultseppunct}\relax
\EndOfBibitem
\bibitem[Müller \emph{et~al.}(2014)Müller, Hönicke, Detlefs, and
  Fleischmann]{M.Mueller2014}
M.~Müller, P.~Hönicke, B.~Detlefs and C.~Fleischmann, \emph{Materials}, 2014,
  \textbf{7(4)}, 3147--3159\relax
\mciteBstWouldAddEndPuncttrue
\mciteSetBstMidEndSepPunct{\mcitedefaultmidpunct}
{\mcitedefaultendpunct}{\mcitedefaultseppunct}\relax
\EndOfBibitem
\bibitem[Parratt(1954)]{Parratt1954}
L.~Parratt, \emph{Phys. Rev.}, 1954, \textbf{95(2)}, 359--369\relax
\mciteBstWouldAddEndPuncttrue
\mciteSetBstMidEndSepPunct{\mcitedefaultmidpunct}
{\mcitedefaultendpunct}{\mcitedefaultseppunct}\relax
\EndOfBibitem
\bibitem[Windt(1998)]{Windt1998}
D.~Windt, \emph{Comput. Phys.}, 1998, \textbf{12(4)}, 360--370\relax
\mciteBstWouldAddEndPuncttrue
\mciteSetBstMidEndSepPunct{\mcitedefaultmidpunct}
{\mcitedefaultendpunct}{\mcitedefaultseppunct}\relax
\EndOfBibitem
\bibitem[Pollakowski and Beckhoff(2015)]{Pollakowski_2015}
B.~Pollakowski and B.~Beckhoff, \emph{Analytical Chemistry}, 2015, \textbf{87},
  7705--7711\relax
\mciteBstWouldAddEndPuncttrue
\mciteSetBstMidEndSepPunct{\mcitedefaultmidpunct}
{\mcitedefaultendpunct}{\mcitedefaultseppunct}\relax
\EndOfBibitem
\bibitem[Reinhardt \emph{et~al.}(2014)Reinhardt, Nowak, Beckhoff, Dousse, and
  Schoengen]{F.Reinhardt2014}
F.~Reinhardt, S.~Nowak, B.~Beckhoff, J.-C. Dousse and M.~Schoengen, \emph{J.
  Anal. At. Spectrom.}, 2014, \textbf{29}, 1778--1784\relax
\mciteBstWouldAddEndPuncttrue
\mciteSetBstMidEndSepPunct{\mcitedefaultmidpunct}
{\mcitedefaultendpunct}{\mcitedefaultseppunct}\relax
\EndOfBibitem
\bibitem[Barth \emph{et~al.}(2017)Barth, Roder, Brodoceanu, Kraus,
  Hammerschmidt, Burger, and Becker]{barth_2017}
C.~Barth, S.~Roder, D.~Brodoceanu, T.~Kraus, M.~Hammerschmidt, S.~Burger and
  C.~Becker, \emph{Applied Physics Letters}, 2017, \textbf{111}, 031111\relax
\mciteBstWouldAddEndPuncttrue
\mciteSetBstMidEndSepPunct{\mcitedefaultmidpunct}
{\mcitedefaultendpunct}{\mcitedefaultseppunct}\relax
\EndOfBibitem
\bibitem[Soltwisch \emph{et~al.}(2017)Soltwisch, Fern{\'{a}}ndez~Herrero,
  Pfl{\"{u}}ger, Haase, Probst, Laubis, Krumrey, and Scholze]{Soltwisch2017}
V.~Soltwisch, A.~Fern{\'{a}}ndez~Herrero, M.~Pfl{\"{u}}ger, A.~Haase,
  J.~Probst, C.~Laubis, M.~Krumrey and F.~Scholze, \emph{Journal of Applied
  Crystallography}, 2017, \textbf{50}, 1524--1532\relax
\mciteBstWouldAddEndPuncttrue
\mciteSetBstMidEndSepPunct{\mcitedefaultmidpunct}
{\mcitedefaultendpunct}{\mcitedefaultseppunct}\relax
\EndOfBibitem
\bibitem[Senf \emph{et~al.}(1998)Senf, Flechsig, Eggenstein, Gudat, Klein,
  Rabus, and Ulm]{F.Senf1998}
F.~Senf, U.~Flechsig, F.~Eggenstein, W.~Gudat, R.~Klein, H.~Rabus and G.~Ulm,
  \emph{J. Synchrotron Rad.}, 1998, \textbf{5}, 780--782\relax
\mciteBstWouldAddEndPuncttrue
\mciteSetBstMidEndSepPunct{\mcitedefaultmidpunct}
{\mcitedefaultendpunct}{\mcitedefaultseppunct}\relax
\EndOfBibitem
\bibitem[Beckhoff \emph{et~al.}(2009)Beckhoff, Gottwald, Klein, Krumrey,
  Müller, Richter, Scholze, Thornagel, and Ulm]{B.Beckhoff2009c}
B.~Beckhoff, A.~Gottwald, R.~Klein, M.~Krumrey, R.~Müller, M.~Richter,
  F.~Scholze, R.~Thornagel and G.~Ulm, \emph{Physica Status Solidi (B)}, 2009,
  \textbf{246}, 1415--1434\relax
\mciteBstWouldAddEndPuncttrue
\mciteSetBstMidEndSepPunct{\mcitedefaultmidpunct}
{\mcitedefaultendpunct}{\mcitedefaultseppunct}\relax
\EndOfBibitem
\bibitem[Lubeck \emph{et~al.}(2013)Lubeck, Beckhoff, Fliegauf, Holfelder,
  Hönicke, Müller, Pollakowski, Reinhardt, and Weser]{J.Lubeck2013}
J.~Lubeck, B.~Beckhoff, R.~Fliegauf, I.~Holfelder, P.~Hönicke, M.~Müller,
  B.~Pollakowski, F.~Reinhardt and J.~Weser, \emph{Rev. Sci. Instrum.}, 2013,
  \textbf{84}, 045106\relax
\mciteBstWouldAddEndPuncttrue
\mciteSetBstMidEndSepPunct{\mcitedefaultmidpunct}
{\mcitedefaultendpunct}{\mcitedefaultseppunct}\relax
\EndOfBibitem
\bibitem[Scholze and Procop(2009)]{F.Scholze2009}
F.~Scholze and M.~Procop, \emph{X-Ray Spectrom.}, 2009, \textbf{38(4)},
  312--321\relax
\mciteBstWouldAddEndPuncttrue
\mciteSetBstMidEndSepPunct{\mcitedefaultmidpunct}
{\mcitedefaultendpunct}{\mcitedefaultseppunct}\relax
\EndOfBibitem
\bibitem[H\"onicke \emph{et~al.}(2016)H\"onicke, Kolbe, Krumrey,
  Unterumsberger, and Beckhoff]{P.Hoenicke2016a}
P.~H\"onicke, M.~Kolbe, M.~Krumrey, R.~Unterumsberger and B.~Beckhoff,
  \emph{Spectrochim. Acta B}, 2016, \textbf{124}, 94--98\relax
\mciteBstWouldAddEndPuncttrue
\mciteSetBstMidEndSepPunct{\mcitedefaultmidpunct}
{\mcitedefaultendpunct}{\mcitedefaultseppunct}\relax
\EndOfBibitem
\bibitem[Pomplun \emph{et~al.}(2007)Pomplun, Burger, Zschiedrich, and
  Schmidt]{pomplun_adaptive_2007}
J.~Pomplun, S.~Burger, L.~Zschiedrich and F.~Schmidt, \emph{Phys. Status Solidi
  B}, 2007, \textbf{244}, 3419--3434\relax
\mciteBstWouldAddEndPuncttrue
\mciteSetBstMidEndSepPunct{\mcitedefaultmidpunct}
{\mcitedefaultendpunct}{\mcitedefaultseppunct}\relax
\EndOfBibitem
\bibitem[Kennedy \emph{et~al.}(1999)Kennedy, Buiu, and Taylor]{Kennedy_1999}
G.~P. Kennedy, O.~Buiu and S.~Taylor, \emph{Journal of Applied Physics}, 1999,
  \textbf{85}, 3319--3326\relax
\mciteBstWouldAddEndPuncttrue
\mciteSetBstMidEndSepPunct{\mcitedefaultmidpunct}
{\mcitedefaultendpunct}{\mcitedefaultseppunct}\relax
\EndOfBibitem
\bibitem[Sherman(1955)]{Sherman1955}
J.~Sherman, \emph{Spectrochim. Acta}, 1955, \textbf{7}, 283--306\relax
\mciteBstWouldAddEndPuncttrue
\mciteSetBstMidEndSepPunct{\mcitedefaultmidpunct}
{\mcitedefaultendpunct}{\mcitedefaultseppunct}\relax
\EndOfBibitem
\bibitem[Schoonjans \emph{et~al.}(2011)Schoonjans, Brunetti, Golosio, del Rio,
  Sol\'{e}, Ferrero, and Vincze]{T.Schoonjans2011}
T.~Schoonjans, A.~Brunetti, B.~Golosio, M.~S. del Rio, V.~Sol\'{e}, C.~Ferrero
  and L.~Vincze, \emph{Spectrochim. Acta B}, 2011, \textbf{66}, 776 --
  784\relax
\mciteBstWouldAddEndPuncttrue
\mciteSetBstMidEndSepPunct{\mcitedefaultmidpunct}
{\mcitedefaultendpunct}{\mcitedefaultseppunct}\relax
\EndOfBibitem
\bibitem[Haase \emph{et~al.}(2016)Haase, Bajt, H{\"{o}}nicke, Soltwisch, and
  Scholze]{Haase2016}
A.~Haase, S.~Bajt, P.~H{\"{o}}nicke, V.~Soltwisch and F.~Scholze, \emph{Journal
  of Applied Crystallography}, 2016, \textbf{49}, 2161--2171\relax
\mciteBstWouldAddEndPuncttrue
\mciteSetBstMidEndSepPunct{\mcitedefaultmidpunct}
{\mcitedefaultendpunct}{\mcitedefaultseppunct}\relax
\EndOfBibitem
\end{mcitethebibliography}
\bibliographystyle{rsc} 

\end{document}